\documentclass[useAMS,usenatbib]{mn2e}
\usepackage{graphics,epsfig}
\usepackage[dvipsnames]{color}

\def \be{\begin{equation}}
\def \ee{\end{equation}}
\def \bea{\begin{eqnarray}}
\def \eea{\end{eqnarray}}
\def\etal{{et al.\ }}

\title[Intracluster gas pressure, entropy injection and redshift evolution]{
Intracluster gas pressure, entropy injection and redshift evolution
}

\author[Biman B. Nath \& Subhabrata Majumdar]
{Biman B. Nath$^1$\thanks{biman@rri.res.in}, Subhabrata Majumdar$^2$\thanks
{subha@tifr.res.in}\\
1. Raman Research Institute, Sadashiva Nagar, Bangalore 560080, India\\
2. Department of Theoretical Physics, Tata Institute of Fundamental
Research, 1 Homi Bhabha Road, Mumbai 400005, India\\
}

\begin{document}


\maketitle

\label{firstpage}

\begin{abstract}
We  study the effect of entropy injection in the intracluster
medium (ICM) 
in light of the recent observationally
 determined universal pressure profile of the 
ICM. Beginning with a power-law entropy profile that is expected in the
absence of any feedback, we show that a simple universal prescription of entropy
injection results in the final, observed universal pressure profile. This
simple prescription has two components, one associated with an overall
increase in entropy and another associated with injection in the central
parts of the cluster. Importantly, both the components of entropy injection are needed
to produce the final universal pressure profile. This is indicative of a need of both 
preheating the ICM as well {\it in situ} AGN/SNe heating.
We demonstrate the usefulness of the method by
extending the calculations to clusters at high redshift, and predict
redshift evolution of cluster scaling relations that can be tested against
data. 
We show that the self-similar
evolution of the universal pressure profile is equivalent to a negative
evolution of entropy injection with redshift, with a scaling
$S_{inj} \propto (1+z)^{-0.8} S_{inj}(z=0)$. We also show
the current observational data is indicative of the entropy injection
decreasing with redshift. 
\end{abstract}


\section{Introduction}
Galaxy clusters are being increasingly used  as cosmological probes. The
evolution of galaxy cluster abundances, for example, is sensitive to 
the values of cosmological parameters, and have been extensively studied
in order to determine or constrain these parameters (Wang \& Steinhardt 1998,
Holder \etal 2001, Majumdar \& Mohr 2003, 2004, Gladders \etal 2007,
Vikhlinin \etal 2009,  Rapetti \etal 2009, 2010,  Rozo \etal 2010,
Sartoris \etal 2010). These attempts 
have been, however, limited by the inadequacy of our knowledge of the state 
of the intracluster medium (ICM) of clusters. Cluster scaling relations
show that the distribution of the ICM gas is different from simple
models of cluster formation in which baryonic gas only reacts to the
gravitational potential of the dark matter (Benson \etal 2004, da Silva
\etal 2004,  Bonaldi \etal 2007, Bonamente \etal 2008).

It is now believed that a number of non-gravitational processes may have
affected the ICM gas, including radiative cooling and energetic feedback
from active galactic nuclei (AGN) and/or supernovae (for example, see
Voit \& Bryan 2001, Nath \& Roychowdhury 2001, Balogh \etal 2001, 
Borgani \etal 2004, Roychowdhury \etal 2004,
Kravtsov \etal 2005, Puchwein \etal 2008,  Battaglia \etal
2010). These processes
tend to raise the entropy of the ICM gas, essentially heating the gas and
pushing it outward, and they make the gas less X-ray bright for a given 
temperature, and help in reconciling theoretical predictions with 
observed data.

In recent years,
the Sunyaev-Zel'dovich (SZ) temperature distortion of the cosmic microwave
background radiation (CMBR) has received a lot of attention, in light of 
the possibility of observing a large number of clusters through SZ effect,
since the magnitude of the effect is independent of redshift (Birkinshaw
1999). The
scaling relations for SZ effect is also altered by the non-gravitational
processes mentioned above, and in general the magnitude of the SZ effect
is decreased by the non-gravitational processes mentioned above (White, Hernquist \& Springel 2002, Roychowdhury, Ruszkowski \& Nath 2005).

Recently, a universal pressure profile for the ICM gas has been shown
to fit the X-ray observations of clusters (Arnaud \etal 2010). This pressure
profile  can be combined with a mass model for the
dark matter halo to yield the corresponding gas density and temperature
profiles, and therefore its entropy profile. In this 
paper, we compare this entropy profile of the ICM (inferred from
the observed universal pressure profile)
to determine the additional entropy needed for the ICM gas, starting with
a physically motivated profile that is expected in the absence of any
feedback. The aim is to determine the
additional entropy needed to explain the observations of local clusters
in a phenomenological manner.

We then speculate upon the possible evolution of entropy injection with
redshift, and calculate the corresponding pressure profiles at high
redshift. 
These profiles can be tested against the data for the evolution of
 cluster scaling relations with redshift
(e.g, between X-ray luminosity and temperature), and the evolution of entropy 
injection with redshift can be inferred with the help of this method. 
We show that although present available data are insufficient to constrain
the evolution of entropy injection, one will be able to iterate with
better data in future and determine the entropy injection at high redshift
from observed cluster scaling relation evolution.

The rest of the paper is structured as follows: In $\S 2$, we show different ways to
calculate the ICM pressure profiles, in $\S 3$ we model injected entropy, in $\S 4$ we look
at its evolution with redshift and we finally conclude in $\S 5$. 
Throughout the paper, we use the subscript $U$ for physical parameters 
related to the universal
pressure profile by Arnaud \etal (2010). Also, we use the subscripts $i$ and $f$ for initial and
final profiles and values.

\section{ICM pressure profile}
Recent observations by Arnaud \etal (2010) have shown that a scaled
universal profile for the pressure of the ICM can fit the X-ray observations.
Defining $x=r/R_{500}$, where $R_{500}$ is the radius within which
the mean over density is $500$ with respect to the critical density,
this profile is given by (eqns 9, 11, 13 of Arnaud \etal 2010),
\be
P_U(x)=P_{500} { P_0 \,  h_{0.7}^2 \over
(c_{500} x)^{\gamma} [1 + (c_{500} x)^{\alpha} ]
^{(\beta-\gamma)/\alpha} } \,
\label{eq:arnaud}
\ee
where,
\bea
P_{500} ({\rm keV} \, {\rm cm}^{-3})&=&1.65 \times 10^{-3} \, E(z)^{8/3} 
 \nonumber\\ && \times
 \Bigl [ {M_{500} \over 3 \times 10^{14} \, h_{0.7}^{-1} \,
M_{\odot} } \Bigr ]^{2/3+ \alpha_p + \alpha_p '(x)} \,, \nonumber\\
P_0&=& 8.403 \, h_{0.7}^{-3/2} \,,
\eea
where $c_{500}=1.177$, $\alpha=1.051$, $\beta=5.4905$, $\gamma=0.3801$,
$\alpha_p=0.12$,
$\alpha_p '(x)=0.1-(\alpha_p+0.1){(x/0.5)^3 \over 1+(x/0.5)^3}$, $M_{500}$ is 
the mass enclosed within $R_{500}$, $H=70 \, h_{0.7}$ km s$^{-1}$ Mpc$^{-1}$,
and $E(z)=H(z)/H_0=[\Omega_0 (1+z)^3+\Lambda_0]^{1/2}$, 
is the ratio of the Hubble constant at redshift $z$ to its 
present value.

We assume a Navarro-Frenk-White (NFW) profile for the dark matter distribution
(Navarro \etal 1997), along with an estimation of the concentration
parameter '$c$' from observations by Comerford \& Natarajan (2007), given
by $c_{vir}={14.5 \pm 6.4 \over 1 +z} (M_{\rm vir}/1.3 \times 10^{13} \, h^{-1} \,
M_{\odot})^{-0.15 \pm 0.13}$. We note that this prescription
refers to the total cluster mass and is an empirical one. 
However, we have found that
our final 
results are largely insensitive to the assumptions of the concentration
parameter (see also Chaudhuri \& Majumdar 2011).

We also use the analytical fit by Bryan
\& Norman (1988) for the over density $\Delta_c (z)=18 \pi^2+82 x-39 x^2$, with
$x=\Omega_m(z)-1$, with which one can calculate the virial radius, as,
\be
r_{vir}= \Bigl ( {M_{\rm vir} \over (4 \pi/3) \rho_{crit} (z) \, \Delta_c(z) } 
\Bigr )^{1/3} \,.
\ee

In order to determine the additional entropy we need to use a default
profile for the ICM gas that is predicted in the absence of any energy
input. Following the tradition in X-ray astronomy, we define
the `entropy', $S$ as $k_B T/n_e^{2/3}$ (which is usually expressed
in the units of keV cm$^2$).

There are three possibilities in this regard: 

\begin{itemize}
\item (a) One can either use
an entropy profile that one expects from the consideration of ICM gas
getting heated at the accretion shock, and in the absence of radiative
cooling. Voit \etal (2005) showed that simulations with only gravitational
processes yield a power law entropy profile of type 
$S_i(r)=1.32 \times S_{200} \,
(r/R_{200})^{1.1}$, where
\be
S_{200}={\mu \mu_e^{2/3} m_p^{5/3} \over 2} \,
 \Bigl [ {2 \pi \over 15} {G^2 M_{200} \over f_b H(z)} 
\Bigr ]^{2/3} \,.
\ee
Here, we use an equivalent expression that uses $R_{500}$, taken from
Pratt \etal (2010), using the observed mean value of $c_{500}=3.2$ for 
relaxed clusters (Pointecouteau \etal 2005),
\be
S(r)/S_{500}= 1.42 (r/R_{500})^{1.1} \,
\ee
where
\be
S_{500}= 106 \, {\rm keV} \rm {\rm cm}^{-2} \,
\Bigl [{M_{500} \over 10^{14}  h_{0.7}^{-1} \, M_{\odot} } \Bigr ]^{2/3}
\Bigl ({1 \over f_g } \Bigr )^{2/3} \, E(z)^{-2/3} h_{0.7}^{-4/3} \,,
\ee
$f_g$ being the baryon fraction. This  assumes
a metallicity of $0.3 \, Z_{\odot}$ and a mean molecular weight $\mu
=0.596$.  This profile appears to fit the
results of simulations without radiative cooling, with approximately
20\% dispersion, especially beyond $r \ge 0.1 R_{200}$ (see also
Mitchell \etal 2009). 

Assuming the initial gas distribution to have this entropy profile,
$S_i(r)$, and assuming this gas to be in hydrostatic equilibrium with
a given total mass profile, the initial gas pressure profile, $P_{g,i}(r)$
 can
be obtained from,
\bea
{d P_{g,i} \over dM_g} 
& = &- {GM(\le r) \over
4 \pi r^4} \nonumber\\
{d r \over dM_g} 
&=& {1 \over 4 \pi r^2 (\mu^{3/5}\mu_e^{2/5}) m_p} \Bigl [ {P_{g, i} \over S_i(r)} 
\Bigr ]^{-3/5} \,,
\label{eq:hydros1}
\eea
where 
$M (\le r)$ is the total mass inside radius $r$,
given by the NFW profile, and $M_g (r)$ is the gas mass inside $r$.
These equations can be solved with the boundary conditions that
$M_g=0$ at $r=0$, and that the total gas fraction inside the virial
radius, $M_g(r_{vir})/M_{\rm vir}=\Omega_b/\Omega_m$.

\item (b)
Simulations that include radiative cooling, however, can also
be used to obtain an initial profile that we can compare with the observed
profile in order to determine the additional entropy needed. Loken
\etal (2002) found from simulations with cooling a universal
temperature profile in the range $0.04<r/r_{vir}<1$, given by,
\be
T_i (r) = 
{1.33 \langle T \rangle \over (1+(1.5r/r_{vir}))^
{1.6}} \,,
\ee
where $\langle T \rangle$ is the emission weighted temperature of the
cluster gas. We 
use an empirically derived scaling relation to determine this
temperature, from the observations of Finoguenov \etal (2001),
(although there may be an order unity difference
between these two definitions of emission weighted temperature)
\be
M_{500}=(2.64^{+0.39}_{-0.34})10^{13} \, {\rm M}
_\odot \Bigl ({ k_b
\langle T \rangle \over 1 \, {\rm keV}} \Bigr )^{1.78}\,.
\ee
We can then compute the initial gas density profile assuming the gas to
be in  hydrostatic equilibrium and using the above mentioned
temperature profile,
\be
{1 \over \rho_{g,i}(r)}{d \over dr} \Bigl [{ \rho_{g,i} k_B T_i \over \mu m_p }
\Bigr ]
 = - {GM(\le r) \over
r^2} \,.
\label{eq:hydros}
\ee
 The gas fraction within the virial radius is assumed to be equal to 
the cosmic baryon fraction, $\Omega_b/\Omega_m$.
The inferred gas density and temperature profiles can then be used
to derive the initial entropy profile 
$S_i(r)=k_b T_i/(\rho_{gas,i}/(\mu_e m_p))^{2/3}$.

\item (c)
One can also use a profile with a polytropic equation of state, $
P \propto \rho ^{\Gamma}$. Li \etal (2010) has recently used this as 
a fiducial profile, with $\Gamma= 1.2$. The normalization comes from the 
assumption of total gas fraction within the virial radius to be same 
as the cosmic gas fraction, as in the above models.

\end{itemize}

The initial pressure profiles resulting from these three possible initial 
entropy profiles 
can be compared with the observed 
 universal pressure profile of Arnaud \etal (2010). 
Given the observed pressure profile, one can get to density, temperature and
entropy profile, by assuming
hydrostatic equilibrium.
 We determine the gas density profile, $\rho _{g,U} (r)$, using
hydrostatic equilibrium condition and the observed universal pressure
(eqn \ref{eq:arnaud}),
\be
{1 \over \rho_{g,U}(r)}{d P_U(r)\over dr}
 = - {GM(\le r) \over
r^2} \,.
\ee
We note that the observed pressure
profile refers only to electron pressure and we should multiply the
electron pressure $P_e=n_ekT_e$, by the ratio of $\mu_e/\mu$, where
$\mu (=0.59)$ is the mean molecular weight, and $\mu_e (=1.14)$ is
the mean molecular weights per free electron, their values being
appropriate for  a gas with metallicity of $0.3 \, Z_{\odot}$. \footnote
{The total pressure, assuming a common temperature, $P=kT\Sigma_i n_i
=kT n_e \, {\Sigma_i n_i \over n_e}=P_e \, {\Sigma_i m_i n_i / n_e
\over \Sigma_i m_i n_i /\Sigma_i n_i}= P_e {\mu_e\over \mu}$.}

Once the gas density profile is calculated, 
the corresponding
temperature profile is then readily computed as $T_U (r)= P_U(r)/(k_b
\rho_{g,U}(r)/(\mu m_p) )$, and then the entropy profile,
$S_U(r)=k_B T_U(r)/(\rho_{g,U}(r)/(\mu_e m_p)^{2/3}$. 

We show in Figure 1 the three initial entropy profiles for a given
cluster mass: the case of universal
temperature profile (dashed line), that of a polytropic equation of state
with $\Gamma=1.2$ (dot-dashed line), and the scaled
entropy profile of Voit \etal (2005) (dotted line). The solid curve shows the
entropy profile corresponding to the observed universal pressure profile  
of Arnaud \etal (2010). As expected, the computed initial profile
using radiative cooling (dashed line) has a higher entropy than the power-law
profile expected without cooling (dotted line). 
It should be noted that these curves should be used only
for $Log_{10} [r/r_{vir}] \ge -1.5$, where the observational data are
more secure than in the inner regions. We can see that the inferred 
entropy at $r \sim 0.1 r_{vir}$ is several times higher than the one 
expected from purely adiabatic case.

\begin{figure}
\centerline{
\epsfxsize=0.4\textwidth
\epsfbox{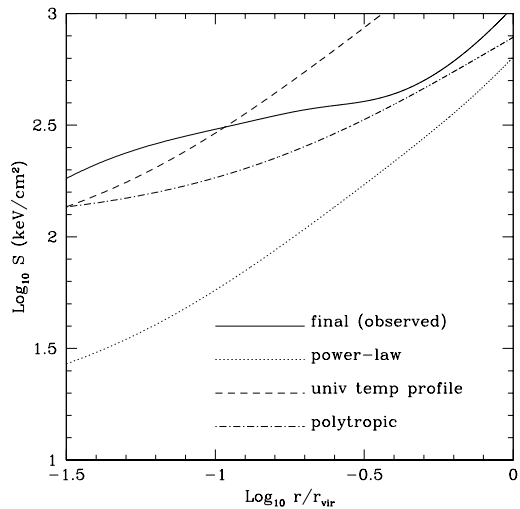}
}
{\vskip-3mm}
\caption{
The entropy profile of ICM gas, for $M_{\rm vir}=10^{14} h^{-1}$ M$_{\odot}$. The
solid line shows the entropy profile corresponding to the observed
universal pressure profile of Arnaud \etal (2010). 
The dotted line shows the scaled power-law
entropy profile from Voit \etal (2005), the dashed line shows the
initial profile computed using the universal
temperature profile of Loken \etal (2002), and the dot-dashed line shows
the entropy profile corresponding to a polytropic gas with $\Gamma=1.2$.
}
\end{figure}

Given these choices for the initial entropy profile, we choose the
one with power-law entropy, of Voit \etal (2005), for the following
reasons. Firstly, the polytropic gas profile lacks physical motivation,
since it is not clear why the ICM gas should  behave in a polytropic
manner in the absence of feedback. Secondly, although the universal
temperature profile from radiative simulations is an appealing choice,
it is not clear as to  what extent it depends on the parameters used in 
the simulations, and also its possible redshift variation is unclear.
On the contrary, the power-law entropy profile is simple and rests on
basic assumptions for the behaviour of gas in a gravitational potential well.

\section{Injected entropy}
Next we compute the additional injected entropy needed to reconcile
the initial profiles with the observed ones.
The extra entropy can be expressed in different ways depending on the
manner in which entropy is injected into the ICM. Previous workers have
characterized the additional entropy as a constant entropy 'floor' that
is independent of the radius. 
For example, Li \etal (2010) have recently 
used a value of $S_{inj}\sim 100\hbox{--}200$ keV
cm$^{-2}$ to determine the ICM pressure profile for their calculation
of the SZ effect. This choice of characterizing the injected entropy
is motivated by the curves showed in Figure 1, in which the gas at
lower radius appears to have an entropy 'floor'.

There is, however, one problem with this characterization. When one
adds entropy to the initial profile, the gas shells move outward. In other
words, entropy added at one position is carried elsewhere by the movement
of the gas, and so the comparison of final and initial entropies at 
a given radial position does not remain physically significant. As Li
\etal (2010) have also recently noted, the injected entropy should be 
characterized by the gas shell in which the injection has occurred. In 
other words, one should compare $S_f(F_g)$ with $S_i(F_g)$, where 
$F_g(r)=M_g(\le r)/M_{\rm vir}$ is the gas fraction inside a radius $r$. 
Note that this gas
fraction is different from the usual definition of gas fraction
in the literature, $f_g=
M_g(r)/M(\le r)$; we distinguish these two distinct parameters by using
two different notations for them.

We show in Figure 2 the corresponding entropy profiles, as functions
of $F_g$, for $M_{\rm vir}=10^{14} h^{-1}$ M$_{\odot}$. 
It is clear that the observed profile (solid line) corresponds
to a case where the gas fraction within $r_{vir}$ has decreased from 
the initial value $\Omega_b/\Omega_m \sim 0.16$, implying outward
movement of gas shells. The entropy 'floor' that appears in the entropy
profile in radial direction exists here only for the very inner shells
of gas. 
For most of the gas in the cluster, the 
(logarithm of) final entropy (solid line) is approximately
separated from the initial
profiles by an additive factor, which translates to a multiplicative
factor for entropy. Viewed in this way, in terms of entropy
injection in Lagrangian
gas shells,
we find it useful to characterize the entropy injection as a 
combination of an 'entropy floor' and a multiplicative factor for
the initial profile:
$S_f(F_g)= S_{c} + \eta S_i(F_g)$.

\begin{figure}
\centerline{
\epsfxsize=0.4\textwidth
\epsfbox{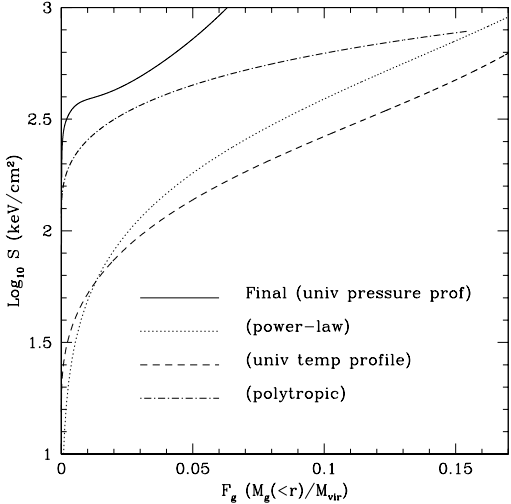}
}
{\vskip-3mm}
\caption{
Same as in Figure 1 except that the profiles are plotted against the
gas fraction $F_g$ instead of the radial coordinate.
}
\end{figure}

The two terms have different physical implications. Note that
the entropy ($S=kT/n_e^{2/3}$) used here is proportional to
the {\it logarithm} of gas entropy. If there had been
no constant term ($S_c$), then we could write the excess energy per
particle as $\Delta \epsilon$, and the entropy $s={\rm Const} \times \ln S$,
 then we have  $\Delta \epsilon=T \Delta s 
\propto T {\Delta S \over S} \propto T (\eta-1)$. Here, since the
temperature is a weak function of radius, a constant $\eta$ would imply
a near uniform injection of energy. 

On the other hand, if the entropy injection were to be only characterized 
with a constant term, then
we would have $\Delta \epsilon \propto T \times 
S_{c}/S_i \propto n_i^{2/3}$, which would imply that energy feedback works
mostly in the inner regions where the gas density is highest. 
Our adopted characterization, with a constant term and a multiplicative
factor essentially allows provisions for 
both these modes of energy feedback.


In order to determine the level of entropy injection (namely, the
value of $\eta$), 
we solve the following equations for hydrostatic equilibrium for the
final gas profile:
\bea
{dr \over dF_g}= &&{M_{\rm vir} \over 4 \pi r^2  \mu m_p} \Bigl ({P_f \over S_f \, k_b} 
\Bigr )^{-3/5} \nonumber\\  
{dP_f \over dF_g} =&& -{G M(\le r)\over r^2} {M_{\rm vir} \over 4 \pi r^2} \,,
\eea
where $F_g=M_g(r)/M_{\rm vir}$. 

The three variables to be determined here are $\eta, P_f(r), F_g(r)$,
and we need three boundary conditions to solve these
equations. One is an obvious choice: (1) $F_g(r=0)=0$. (2) 
Another obvious choice is that $P_f (R_{500})= P_U(R_{500})$, so 
that the final pressure profile corresponds to the observed universal
pressure profile. (3) Lastly we use the value of
gas fraction at an outer radius, but this requires an explanation.

For the third boundary condition, we use the value of $F_g(R_{500})$
(i.e., $M_g(R_{500})/M_{\rm vir}$) that
corresponds to the universal pressure profile. 
We note that the gas
fraction at $R_{500}$ as given by the universal pressure profile is
different from the ratio of $\Omega_b/\Omega_m\sim 0.167$, which implies
that a fraction of the original ICM gas has been thrown out of the system,
if one considers the ratio $\Omega_b/\Omega_m$ to be universal initially.
We show in Figure 3 the gas fraction implied by the universal pressure
profile, and compare with other observations. Note that
observations of gas fraction within $R_{500}$ usually refers to
$f_g (R_{500})=F_g(R_{500}) \times ({ M_{\rm vir} \over M (<R_{500})})$. 
In Figure 3 we show the values of $f_g (R_{500})$ as given by the
universal pressure profile (solid line) for clusters at $z=0$ and compare
them with the observed values (dotted line), as reported by
Sun \etal (2009). 

We also show by the dashed line another ratio, that of the gas mass
within a radius $2 R_{vir}$ and the total virial mass of the cluster.
This ratio, $F_g(2 R_{vir})={ M_g ( 2 R_{vir}) \over M_{vir}}$, is found to 
be {\it almost a constant} for all clusters, and close to the value of
$\Omega_b/\Omega_m$, shown by the dot-dashed line, with an average
deviation of order $\sim 10\%$. This interesting
result suggests that a fraction of the ICM gas, originally inside $R_{vir}$
and with a universal gas fraction of $\Omega_b/\Omega_m$, gets ejected
due to the feedback process, but that the gas does not extend beyond
a radius $2 R_{vir}$. 

Entropy feedback predominantly affects the gas while keeping the dark matter unchanged.  In this case, gas mass conservation will 
ensure that $M_g (R_{vir}, i)=M_g(2 R_{vir}, f)$, where the latter term
is the gas mass within $2 R_{vir}$ after the feedback processes. Dividing
both sides by $M_{vir}$, and recalling that the initial profile is normalized
by $M_g(R_{vir},i)/M_{vir}=\Omega_b/\Omega_m$, yields the condition for
the final profile, 
$M_g(2 R_{vir})/ M_{vir}=\Omega_b/\Omega_m$. The choice of the final radius to be
$2 R_{vir}$ is slightly arbitrary. However, this choice 
 is consistent with the findings of Battaglia \etal
2010, who show that there is significant contribution to the 
SZ flux from beyond the
virial radius, and going beyond $2 R_{vir}$ does not add to the SZ power.
 This result  allows us to use a boundary
condition, that the gas fraction $F_g(2 R_{vir})={M_g (2 R_{vir}) \over M_{\rm vir}}
\sim \Omega_b/\Omega_m$.
We also show in Figure 3 by two thin dashed lines the results for $F_g (2.5 R_{vir})$
and $F_g (1.5 R_{vir})$, and the total spread is of order $17\%$ of the values of
$F_g (2 R_{vir})$. This is the level of uncertainty in our results that stems from the
lack of precise knowledge about the outer boundary condition.

\begin{figure}
\centerline{
\epsfxsize=0.4\textwidth
\epsfbox{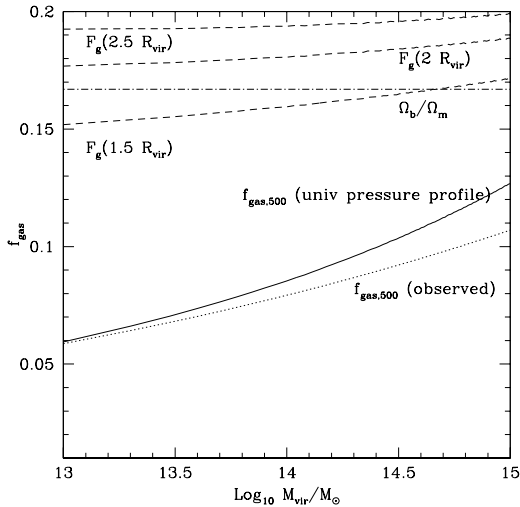}
}
{\vskip-3mm}
\caption{
Gas fraction within $R_{500}$ ($f_g=M_g(R_{500})/M(R_{500}))$ is plotted
as a function of cluster virial mass, for the universal pressure profile
(solid curves), for $z=0$. The dotted line shows the observations
of Sun \etal (2009) and the dot-dashed line shows the value of $\Omega_b/
\Omega_m$.  The topmost (solid) dashed line shows the fraction $F_g (2 R_{vir})$,
which is the ratio $M_g (2 R_{vir})/M_{vir}$, as computed from extending
the universal pressure profile to $2 R_{vir}$. Interestingly, this line is
close to the Universal gas fraction to within $10-15\%$ for all masses.
The pair of thin dashed lines above and below this curve are for $F_g (2.5 R_{vir})$
and $F_g (1.5 R_{vir})$, respectively. 
}
\end{figure}

We find that with the above mentioned assumptions and boundary conditions,
the following prescription of entropy injection works satisfactorily
in converting the initial entropy profile to the final profile as given
by the universal pressure profile at $z=0$, 
for $10^{14} \le M_{\rm vir}/M_{\odot} \le 10^{15}$:
\be
S_{inj}(F_g)= 200 \, {\rm keV} \, {\rm cm}^{-2} + 2 
\times  (\frac{M_{\rm vir}}{10^{14} \, h^{-1} \, 
M_{\odot}})^{-0.2} \times S_i(F_g) \,.
\label{eq:pres}
\ee
 
The values of the coefficients provide the best match within an accuracy
of $\sim 15\%$.
This form of the feedback implies that low mass clusters need overall 
more feedback than rich clusters,
although (from the constant term) all clusters appear to need feedback in
their central regions.

Physically these two modes may refer to preheating and
AGN/SN feedback processes, one in which all the gas particles in the ICM
are energized and another, in which particles in the central region get
more entropy than elsewhere.
We show below that from {\it the comparison of the final pressure profile
and the initial pressure profiles, that it would be impossible to achieve the
final pressure profile with only a single mode of entropy injection}.

We show in Figures 4 and 5 two examples of the result of this type of entropy
injection, for $M_{\rm vir}=10^{15} h^{-1}$ M$_{\odot}$ and 
$M_{\rm vir}=10^{14} h^{-1}$ M$_{\odot}$, respectively, at z=0. 
The top-left panels of both figures show
the density profiles: dashed lines show the initial profile of power-law
entropy; the dot-dashed lines show the profile corresponding to the 
universal pressure profile and the dotted lines show the final profile, which
is a result of the entropy injection according to the above mentioned
prescription.  The top-right panels show the temperature profile, the 
bottom-left panels show the pressure profile and the bottom-right panels
show the entropy profiles.

Before discussing the results, let us summarize the method of our calculation.
We have began with a power-law entropy profile, and used a prescription
for entropy injection, which along with the above mentioned boundary conditions
yield a pressure profile. 
We iterate a loop until the three boundary conditions are matched (within
$10\%$). This leads to a match with the observed universal pressure profile
because of the chosen entropy injection prescription (\ref{eq:pres}).
As a result, we find in Figures 4 and 5 that the final pressure
profiles (bottom-left panels, dotted lines) match well the observed profile
(dot-dashed lines). 
The match for the temperature
(top right panels)
and entropy (bottom right panels) profiles are less satisfactory, but
we note that the dot-dashed lines here are not observed profiles, but
inferred from the universal pressure profiles. Since we have decided to
peg our boundary conditions on the observed pressure profiles, we 
conclude that the above entropy injection prescription gives a
satisfactory match with observations. {\it Since a single prescription
appears to work for the whole range of cluster masses, one could call it
a universal entropy injection prescription}.

\begin{figure}
\centerline{
\epsfxsize=0.4\textwidth
\epsfbox{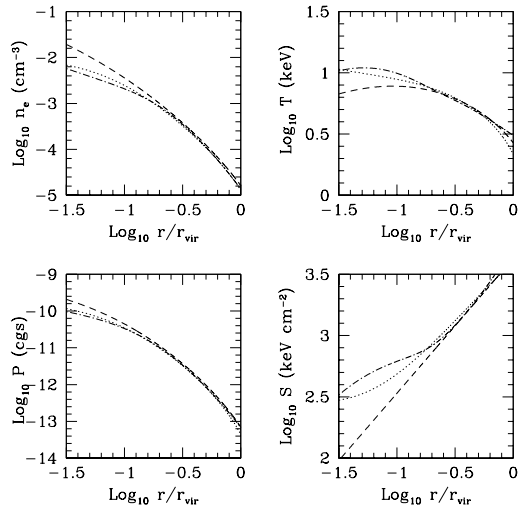}
}
{\vskip-3mm}
\caption{
Comparison between the initial and final profiles, before (dashed) and after
(dotted) entropy injection, and those obtained from universal pressure profile
(dot-dashed), for $M_{\rm vir}=10^{15} h^{-1}$ M$_{\odot}$. The top-left panel
shows the density profiles, the top-right panel shows the temperature
profiles, the bottom-left panel shows the pressure profiles and the
bottom-right panel shows the entropy profiles. 
}
\end{figure}

\begin{figure}
\centerline{
\epsfxsize=0.4\textwidth
\epsfbox{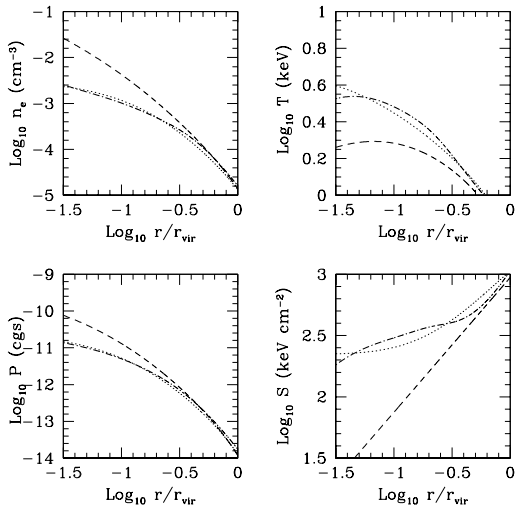}
}
{\vskip-3mm}
\caption{Same as Figure 4a, but for $M_{\rm vir}=10^{14} h^{-1}$ M$_{\odot}$.
}
\end{figure}

\section{Evolution with redshift}
\subsection{Evolution of entropy and pressure}
Our exercise has so far been limited to observed clusters at low 
redshifts: we have used a physically motivated initial gas profile,
and compared it to the observed profile to deduce a prescription
for entropy injection. It is true that the entropy injection
prescription allows us to speculate on the kind of feedback process that may
give rise to the entropy enhancement. The prescription, however,
remains a phenomenological one, and we wish to explore its usefulness
beyond the regime within which it has been deduced 
in the previous section.

In particular,
we wish to study the ICM in high redshift galaxy clusters, and 
determine the amount of feedback that may have occurred at high redshift.
The properties of the ICM at high redshift have been studied through
the observations of evolution of cluster scaling laws, e.g., the
relations between X-ray luminosity and temperature, and with the help
of numerical simulations. In the rest of the paper, 
we show that our semi-analytical
formalism can help in this regard, by determining the amount of 
feedback needed to produce an ICM profile that is consistent with
redshift evolution of cluster scaling relations. Voit (2005) had
suggested that cluster scaling relations would differ at high redshifts
owing to the effects of feedback and radiative cooling (see also
Muanwong, Kay \& Thomas 2006; Short \etal 2010). 

At present, however, 
the usefulness of this
method is limited by the lack of accurate determinations of
scaling relation evolutions. Ettori \etal (2004) found that the
evolution of $M-T$ relation in a sample of 28 clusters
at $0.4 < z < 1.3$ followed the self-similar model, expected from
simple cosmological considerations. But they found the normalization
of the $L_x-T$ relation to decrease with redshift, which was later
confirmed by Hilton \etal (2007) from a study of a high redshift
cluster with $XMM-Newton$ (see also Branchesi \etal 2007). 
Morandi \etal (2007) found negative
evolution in the $M-T$ relation but a positive evolution for $L_x-T$ 
relation (see also Maughan \etal 2007). At the same time, Maughan \etal (2006)
found little evidence for any such redshift evolution apart from
that expected from self-similarity.

One reason for this uncertainty is the scatter produced by the large X-ray
luminosity at the centres of cool core clusters (Chen \etal 2007). Cluster
mergers also produce deviations (Kravtsov \etal 2006). Also, the
sample of small number of clusters at high redshift suffers from
selection effects (Pacaud 2007). It is hoped that a large and homogeneous
sample will be gleaned from the upcoming XMM Cluster Survey (XCS)
(Romer \etal 2001), and allow a better determination of the
evolution of X-ray scaling relations.

\begin{figure}
\centerline{
\epsfxsize=0.4\textwidth
\epsfbox{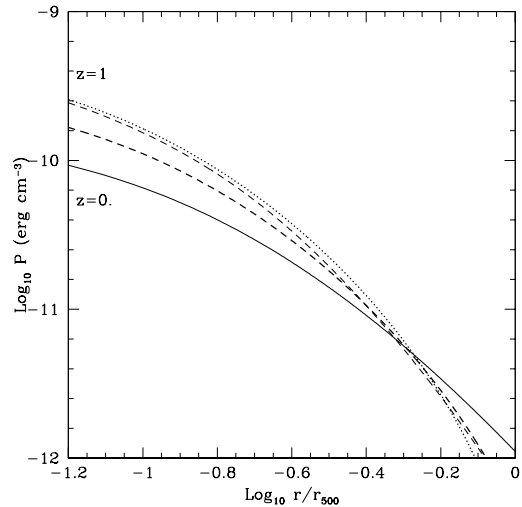}
}
{\vskip-3mm}
\caption{
The pressure profiles for $M_{\rm vir}=10^{15} h^{-1}$ M$_{\odot}$ at $z=0,$
and $1$
are plotted against $r/R_{500}$ (in cgs units), for the case of 
constant entropy enhancement. The solid line refers
to the $z=0$ profile,  and the thick dashed line is for $z=1$.
The thin dashed line show the universal pressure profile scaled
by a factor $E(z)^{8/3}$ for comparison. We also show with dotted line
the case for the entropy injection scaling as $(1+z)^{-0.8}$, which
matches closely the self-similarly scaled universal pressure profile.  
}
\end{figure}

Keeping these uncertainties in the observed relations in mind, we 
would like to extend our phenomenological study to high redshift clusters.
For this exercise, we need to assume a redshift
evolution of the feedback prescription. Also, instead of changing the
boundary conditions with redshift, we assume that the boundary conditions
at high redshift is the same as at $z=0$, namely, $F_g(r=0)=0$, and that 
$F_g(2 R_{vir})=\Omega_b/\Omega_m$. This will yield the gas profiles,
which can then be compared with the
expectations from scaling relation evolutions. This process can then
be iterated
 to determine the amount of feedback that is required to 
explain the evolution of the cluster scaling relations. This 
iteration
will make sense with accurate determination of scaling relation evolutions,
and in the absence of such observations, we only demonstrate the first
part of this process, namely, prediction of ICM at high z corresponding
to a given feedback prescription and how this can be used to compare with
observed cluster scaling relation evolutions.


We have used simple modifications to the entropy injection prescription. Here,
one can work out a number of possibilities, parameterized by a certain
set of variables and test the resulting pressure profiles against data. In
this paper, we demonstrate the utility of our approach by calculating the
cluster scaling relations with redshift resulting from a simple redshift
scaling of the entropy injection prescription, namely,
$S_{inj} (z)= (1+z)^\alpha S_{inj, z=0}$. This means that we multiply the
expression of injected entropy in eqn \ref{eq:pres} by a factor $(1+z)^\alpha$.

Assuming this feedback prescriptions, we calculate the ICM profiles
for clusters at high redshift for different values
of the power-law index $\alpha$. 
In Figure 6, we show the resulting pressure 
profiles for $\alpha=0$, for
$M_{\rm vir}=10^{15} h^{-1}$ M$_{\odot}$ at $z=0$ (thick 
solid line)and $1$ (thick dashed line).
The thin line 
shows the self-similarly scaled universal
pressure profile at $z=1$, scaled by the factor $E(z)^{8/3}$. 
The resulting profiles clearly  have larger values for pressure than
those expected from a self-similar evolution of the pressure profile.
We can similarly calculate the pressure profiles for other cluster masses
and at different redshifts.

We also show in Figure 6 the case of $\alpha=-0.8$, for a negative
evolution of entropy injection, at $z=1$ and we find that this matches the
self-similarly scaled universal pressure profile closely. We have checked
this at other redshifts $z<1$ and found that they match within an accuracy
of $25 \%$.
Our calculations with different values of the parameter $\alpha$ show
that the pressure profiles at high redshift are increasingly
suppressed with increasing
values of $\alpha$, as expected. This will result in different 
forms of evolution of various cluster scaling relations, which can be
tested against data, and we now turn to a discussion on this issue.

\subsection{Evolution of scaling relations}
How do these profiles
compare with the observations of cluster scaling relations? Following
Short \etal (2010), we study the normalizations of the $L_x-T$ and 
the SZ decrement $Y_x-M$ relation (see below) and their evolution in redshift.
For the $L_x-T$ relation, we assume a power-law relation,
\be
L_x \, E(z)^{-1}=C1 (T_{sl}/6 keV)^\alpha \,,
\ee
where $L_x$ is the X-ray luminosity at $0.5-1$ keV band and $T_{sl}$
is the spectroscopic like temperature, as defined
by Mazzotta \etal (2004). One can study the evolution
of the normalization $C1$, with redshift, for ICM profiles given by
self-similarly scaled universal pressure profile, and those with
other forms of evolution.

\begin{figure}
\centerline{
\epsfxsize=0.4\textwidth
\epsfbox{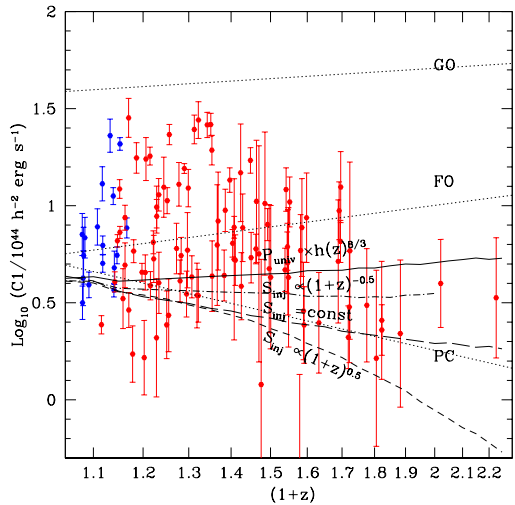}
}
{\vskip-3mm}
\caption{
The evolution of the normalization $C1$ of $L_x-T$ relation is shown
with redshift, for universal pressure profile with self-similar
scaling (solid line) and three cases of enhanced pressure profile:  for 
a constant entropy injection at all redshifts (long dashed line),
for $S_{inj}\propto (1+z)^{-0.5}$ (dot-dashed line), and 
$S_{inj}\propto (1+z)^{0.5}$ (short dashed line).
The three dotted lines show
the results of simulations GO, FO and PC (see text for details).
}
\end{figure}

In Figure 7 we show the redshift evolution of $C1$, in the units of
$10^{44} h^{-2}$ erg s$^{-1}$. The solid line shows the evolution
obtained from a self-similarly scaled universal pressure profile,
which means that the profile is scaled by a factor $E(z)^{8/3}$. 
The
long dashed line shows the case
of entropy injection being constant with redshift, 
the dot-dashed line shows the case of the scaling $(1+z)^{-0.5}$,
whereas the short-dashed line shows the results of $S_{inj}\propto
(1+z)^{0.5}$.

This figure can be compared with Figure
12 of Short \etal (2010), which shows the results of different 
simulations and observations. As mentioned earlier, the data
has large scatter and the form of evolution remains unclear. 
Simulations
with feedback, however, show either positive evolution 
(from FO simulation including AGN and supernova feedback,
 by Short \etal (2010)) or negative evolution
(PC simulation, including preheating, by Hartley \etal 2007).
Simulations without any feedback (GO, in the Millennium Gas simulations
series; see Short \etal 2010) show a mild positive evolution.
However, the amplitude without feedback fails to match observations. 
Thus some form of energy injection is absolutely necessary.

We show in Figure 8 the redshift evolution of the normalization
$C2$ of the relation between $Y_{x}-M_{500}$ relation,
normalizing at $M=5 \times 10^{14} h^{-1}$ M$_{\odot}$,
\be
Y_{\rm X} \, E(z)^{-2/3} =C2 \Bigl ({M_{500} \over 5 \times 10^{14}\,  
h^{-1} \,  M_{\odot}} \Bigr )^\alpha \,.
\ee 
It has been recently shown that $Y_x=M_g T$ (evaluated within a certain
over density, here $R_{500}$) behaves as a low-scatter proxy for the
cluster mass, regardless of the dynamical state of the cluster
(e.g., Kravtsov \etal 2006), since it essentially expresses the total
thermal energy of the gas which is not strongly affected by
cluster mergers. 

The parameter
$C2$ has the units of $keV M_{\odot}$. It has been shown that the universal
pressure profile is consistent with SZ observations at low redshift
(Arnaud \etal 2010). We therefore do not show the scaling law at $z=0$,
but focus on the evolution of the normalization with redshift.
Again, the curves in Figure 8
shows the expectations from a self-similarly scaled universal pressure
profile (solid line) and  those from entropy injection at high
redshift with the following scalings: $\propto (1+z)^{0.5}$ (short-dashed line),
constant in redshift (long dashed line), and $\propto (1+z)^{-0.5}$ (dot-dashed line).
The dotted lines again show the results of three types
of simulations mentioned above, without feedback (GO), with AGN/supernova
feedback (FO) and with pre-heating (PC), as reported in Short \etal (2010).

\begin{figure} 
\centerline{
\epsfxsize=0.4\textwidth
\epsfbox{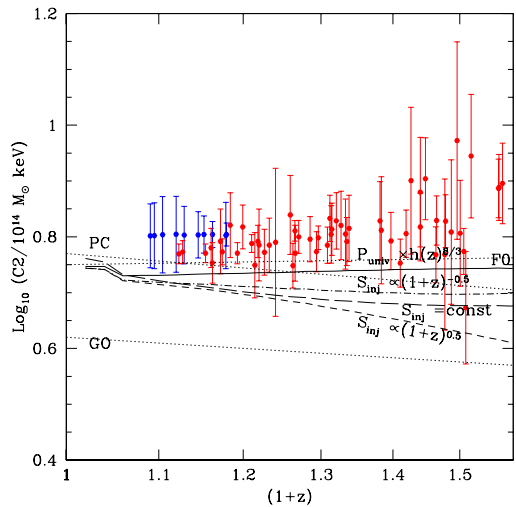}
}
{\vskip-3mm}
\caption{
Same as Figure 7, but for the normalization parameter $C2$
of $Y_{\rm X}-M_{500}$ relation. Data points from 
REXCESS (Pratt \etal
2009) are shown as blue and high redshift
data from Maughan \etal (2006) are shown with red points. 
}
\end{figure} 

\section{Discussion}
Figure 7 shows that 
the self-similar evolution of the universal pressure profile (solid line)
is consistent
with the available data. The case for constant entropy injection
(dot-dashed line) and positive evolution do not fare well, 
and the match with the data 
improves for negative redshift evolution (dashed line) of entropy injection. 
As mentioned earlier, the self-similarly scaled universal pressure profile
is equivalent to a scaling of entropy injection as $(1+z)^{-0.8}$, and the
solid line therefore also shows the case of a steep negative evolution
with $\alpha=-0.8$.
The results from enhanced entropy are
largely consistent with the range of results from current simulations
including feedback and preheating (shown by the dotted lines marked as
FO and PC). 

The curves in figure 8 show that 
the self-similar evolution of the universal pressure profile is consistent
with the available data, and that the data favour negative evolution
of entropy injection in general. The 
enhanced profile with our simple scaling law are 
consistent with the results from simulations including feedback.
Comparing with the data, which include 
low-redshift observational data from REXCESS (Pratt \etal
2009; Short \etal 2010) and high redshift
data from Maughan \etal (2006), we find that the
case of constant entropy injection matches better and, as in the case
of $L_x-T$ relation, the situation improves with negative redshift
evolution. 
A more detailed analysis of the likelihood of the values of $\alpha$
that is allowed by the data is beyond the scope of the present paper, 
and we will pursue this approach in a future paper, and determine
the possible evolution of entropy injection, parameterized by certain variables
whose values could be determined by comparing with existing data.
But the trend shown in Figure 8, indicates that 
the available data
require an evolution of the
entropy injection that scales as $(1+z)^{\alpha}$ with $\alpha \le  -1$. 
Note that owing to the non-gravitational nature of the feedback processes,
the enhanced entropy need not evolve self-similarly with
redshift. However, the entropy of gas as determined by gravity alone
is expected to evolve self-similarly, namely, $S \propto E(z)^{-2/3}$
for clusters with same mass.
Nevertheless, we can compare the entropy scaling inferred above with 
the self-similar scaling. In $\Lambda$-CDM model, $E(z)^{-2/3}$ roughly
scales as $(1+z)^{-0.4}$ in this redshift range. Therefore, the
scaling $S_{inj}
\propto (1+z)^{-1}$ can be written as $S_{inj} \propto E(z)^{-2/3} (1+z)^{0.6}$,which is indicative of a non-standard evolution.

A more important conclusion that we can draw from the previous
section, however, is that that in the presence of accurately
determined cluster scaling relation evolution, as is expected in the
near future, one will be able to
compare the predictions from entropy injection with that from redshift
evolution of cluster scaling relations, and therefore iterate
in order
to determine the entropy enhancement at high redshift. 

Fang \& Haiman (2008) have argued, on the basis of
a simple model of evolving
entropy floors for the ICM with redshift and comparing the results
with observed evolution of scaling relations, that the entropy floor
at high redshift is likely to be shallower than that at present.
They argued that a deceasing entropy floor with redshift is expected
from feedback processes due to active galactic nuclei (AGN) and supernovae
driven winds. 

It is however difficult to interpret the amount of 
feedback only from the level of the entropy floor, since the
default entropy at high redshift is also likely to be small, from
cosmological evolution.
Entropy, as defined by $S\equiv T/n_e^{2/3}$, would normally
evolve as $E(z)^{-2/3}$, for clusters with same virial mass,
in the absence of any feedback process. Any deviation from this 
can be identified as the result of a feedback process. For example,
Ettori \etal (2004) found that for clusters with same temperatures,
the entropy floor evolves as $E(z)^{4/3} \, S \propto (1+z)^{0.3}$.
Nath (2004) interpreted this as the result of increasing feedback
energy deposition at higher redshifts, possibly as a result of
increase in the abundance of AGNs at high redshift.

Our work in this paper presents a method of determining the amount of
feedback and its evolution with redshift, by using an initial
entropy profile and injecting entropy into the profile (defined
by radially distributed gas shells). It remains to be seen how the
entropy injection prescription is related to any physical process
of feedback, and we will return to this issue in a future paper.

Finally, we note that our results are accurate up to $\sim 15\%$, given the
accuracy with which we have implemented the boundary conditions in 
arriving at our results. Our analysis presented here has ignored
the contribution from non-thermal pressure of the gas, which can
lead to underestimating the mass by $15-20\%$
around $r_{2500}$ at $\sim 25\%$ at the virial radius (see for example, Rasia
\etal 2004), but which would 
have introduced additional free parameters, and more uncertainties
in the interpretation of cluster scaling relation data,  the
results of the entropy injection prescription. The simplicity of the
entropy injection prescription presented in this paper, however,
may make it possible to include non-thermal pressure.

There is another source of uncertainty that arises from the fact that
a fraction of the baryons in clusters would form stars, and would thus
affect the boundary condition for normalization. Although Gonzalez \etal
(2007) found that the stellar mass fraction in a cluster of $M_{500}
=10^{14}$ M$_{\odot}$ can be as large as $0.25 \Omega_b/\Omega_m$
(see also Giodini \etal 2009; Andreon \etal 2010), Balogh
\etal (2008) have found that such a large fraction, if confirmed, would
cause problem for the paradigm of hierarchical structure formation,
let alone being an 'overcooling' problem. On the other hand a conservative
estimate from Bower \etal (2006) is that the stellar fraction is 
$\sim 0.1$ times the cosmic baryon fraction. 
This is also in agreement with hydro simulations of
clusters (Ettori \etal 2006). Recently, Simionescu \etal (2011) have found
from Suzaku observations of the Perseus cluster that the baryon fraction
approaches the Universal value much earlier than the virial radius. All of these would imply that our
boundary conditions used here would be modified by a factor $\sim (1-0.1)$,
and the resulting profiles would differ by $\sim 10\%$ and, hence, would
decrease the required entropy injection by a similar fraction.
We have neglected the contribution of star formation in this paper 
because our entropy injection prescriptions
are accurate up to $\sim 15\%$.

\bigskip
We would like to thank the referee for a very careful reading of the manuscript that has helped us in 
improving the paper. We thank Satej Khedekar for help with Figures 7 and 8.

\end{document}